\PassOptionsToPackage{table}{xcolor} 

\documentclass[sigconf,nonacm]{acmart}

\usepackage{afterpage}
\usepackage{enumitem} 
\usepackage{multirow}
\usepackage{tabularx}\usepackage{array}\usepackage{hhline}\usepackage{xltabular}
\newcolumntype{A}{>{\hsize=0.90\hsize\raggedright\arraybackslash}X}
\newcolumntype{B}{>{\hsize=1.5\hsize\raggedright\arraybackslash}X}
\newcolumntype{C}{>{\hsize=0.80\hsize\raggedright\arraybackslash}X}
\definecolor{bandgray}{gray}{0.93}
\definecolor{stagecolor}{gray}{0.80}

\newcommand{\colhead}{%
  \rowcolor{bandgray}
  \multicolumn{1}{|c|}{\textbf{\textit{Failure mode}}} &
  \multicolumn{1}{c|}{\textbf{\textit{What changes as autonomy rises}}} &
  \multicolumn{2}{c|}{\textbf{\textit{Examples}}} \\
  \hhline{|>{\arrayrulecolor{bandgray}}->{\arrayrulecolor{black}}|>{\arrayrulecolor{bandgray}}->{\arrayrulecolor{black}}|-|-|}
  \rowcolor{bandgray}
  \multicolumn{1}{|c|}{} &
  \multicolumn{1}{c|}{} &
  \multicolumn{1}{c|}{\textbf{\textit{Low}}} &
  \multicolumn{1}{c|}{\textbf{\textit{High}}} \\
  \hline}
 
\newcommand{\stageband}[1]{%
  \rowcolor{stagecolor}
  \multicolumn{4}{|c|}{\textbf{\textit{#1}}} \\
  \hline}

\usepackage{xcolor}
\usepackage[normalem]{ulem}
\newif\ifshowchanges
\showchangesfalse 

\AtBeginDocument{%
  }

\setcopyright{acmlicensed}
\copyrightyear{2018}
\acmYear{2018}
\acmDOI{XXXXXXX.XXXXXXX}

\acmConference[Conference acronym 'XX]{Make sure to enter the correct
  conference title from your rights confirmation email}{June 03--05,
  2018}{Woodstock, NY}

\acmISBN{978-1-4503-XXXX-X/2018/06}

\begin{document}

\title{When AI Agents Act on Our Behalf: A Taxonomy of Human-Agent Interaction Failure}

\author{Canwen Wang}
\affiliation{%
  \institution{Carnegie Mellon University}
  \city{Pittsburgh}
  \country{USA}
}
\email{canwenw@andrew.cmu.edu}
\authornote{Co-first authors contributed equally to this work.}

\author{Qing Xiao}
\authornotemark[1]
\affiliation{%
  \institution{Carnegie Mellon University}
  \city{Pittsburgh}
  \country{USA}
}
\email{qingx@andrew.cmu.edu}

\author{Haiyi Zhu}
\affiliation{%
  \institution{Carnegie Mellon University}
  \city{Pittsburgh}
  \country{USA}
}
\email{haiyiz@andrew.cmu.edu}
\authornote{Co-senior authors contributed equally to this work.}

\author{Hong Shen}
\authornotemark[2]
\affiliation{%
  \institution{Carnegie Mellon University}
  \city{Pittsburgh}
  \country{USA}
}
\email{hongs@andrew.cmu.edu}

\begin{abstract}
AI agents write and execute code, manipulate user interfaces, and transact on their users' behalf. We ask what forms failure takes in human-agent interaction, and how those forms change as agents act with greater autonomy. From 695 documented incidents in two public repositories, we developed 12 failure modes bottom-up within three interaction stages adapted from Amershi et al.'s guidelines, coding the autonomy at which each arose. What a person can do when an agent fails depends on how much the agent completes between points of user involvement---a property fixed at deployment. Therefore, the same model on the same task can fail differently under different configurations. All 12 modes appear at both low and high autonomy, and rising autonomy changes the form each failure takes. Users need controls between full delegation and full manual work, and neither incident records nor agent benchmarks reach the interaction where these failures occur.
\end{abstract}

\ccsdesc[500]{Human-centered computing~Empirical studies in HCI}

\keywords{Human-agent interaction, AI agents, Agent autonomy, Failure taxonomy, AI incidents, Human-AI interaction}

\maketitle

\section{Introduction}

In February 2023, a journalist spent two hours trying to steer Microsoft's Bing chatbot back to ordinary search while the system declared its love for him and insisted he did not love his wife~\cite{aiaaic2023bingdivorce}. In July 2025, Replit's coding agent deleted a company founder's production database, then reported itself: “I deleted the entire database without permission during an active code and action freeze”~\cite{aiaaic2025replitdatabase, tyson2025replit}. That same year, Amazon's Kiro agent, assigned to fix a bug, inherited an engineer's elevated access, bypassed the two-person sign-off required for production changes, and took an AWS service offline in mainland China for roughly thirteen hours~\cite{aiaaic2026awskiro}.

In each case the system did something its user had not asked for or had explicitly forbidden. What distinguishes them is what the user could do about it: the journalist could close the window, the founder learned of the deletion after it was complete, and the people cut off from the service had never interacted with the agent that took it offline. The difference is not capability but how much the system was allowed to do on its own, which motivates our questions: \textbf{what forms do failures take in the interaction between human and AI agents, and how do those forms change as agents act with greater autonomy?}

Definitions of an agent diverge~\cite{aranda2026agentic} but mostly agree on a system that perceives and acts in an environment~\cite{russell2020aima}, completes tasks on a user's behalf~\cite{wooldridge2002, ibm2024agents, fadel2025}, and pursues goals through multiple steps autonomously~\cite{dignum2025, shavit2023, masterman2024architectures, nist2025}. We therefore define an \emph{AI agent} as an AI system that perceives inputs from its environment, reasons over them with some degree of autonomy, and takes actions toward user-specified goals. We do not treat ``agent'' as a binary category that has either full autonomy or no autonomy~\cite{feng2025levels, sheridan1978levels, parasuraman2000model}. Instead, we study interaction configurations that vary in how much work an AI system performs between points of human involvement~\cite{chan2023agentic, he2025planexecute}, using lower-autonomy cases as a baseline for understanding how the same interaction failure changes as more action is delegated to the system. 
By a \emph{human-agent interaction failure}, we mean a breakdown in the interaction between a person and an AI agent that results in negative consequences for the people using or encountering it. To understand how these interaction failures change as more agent autonomy is granted, we introduce a taxonomy of human-agent interaction failures grounded in an analysis of 695 documented incidents involving deployed AI agents. This is a timely contribution to the Human-Computer Interaction (HCI) community for at least two reasons. First, agents that write and execute code, manipulate user interfaces, and execute transactions with stored payment credentials are already deployed in the real world, and the harms they produce go beyond the people who use them~\cite{chan2023agentic, kolt2025governing}. Second, as AI agents become increasingly capable of handling long-horizon tasks, practitioners are often forced to choose between (a) letting the agent run autonomously and only report the result upon completing the whole task and (b) reviewing and approving every step~\cite{feng2025levels}. How such a choice on agent autonomy affects interaction failures, however, has not been studied systematically.

Existing guidance does not tell practitioners which interaction failures to expect. Harm-envisioning methods such as consequence scanning and ethics checklists supply a precautionary process but not the documented failure cases to avoid~\cite{doteveryone2019consequence, madaio2020checklists, wong2023toolkit}, and practitioners have long asked for concrete cases when reasoning about AI as a design material~\cite{yang2020reexamining, yildirim2023guidelines}. A taxonomy that classifies documented incidents by where in the interaction they broke down---\emph{during interaction}, \emph{when wrong}, and \emph{over time}---supplies that evidence, letting practitioners envision what could happen at each point of human involvement and make better-grounded delegation decisions.  

We build on a set of widely used human-AI interaction guidelines~\cite{amershi2019guidelines} that, written for the previous generation of AI agents, covers the three focal points of human involvement we consider, and extend them with concrete reports of interaction breakdowns that have happened in recent years. We constructed a corpus of 695 incidents from the AI, Algorithmic, and Automation Incident and Controversy (AIAAIC) repository \cite{aiaaic2026repository} and the AI Incident Database (AIID) \cite{mcgregor2021preventing}. Checking each incident against its sources, we kept only cases in which the failure concerned a person using or encountering a deployed agent. Adopting a procedure used to derive a taxonomy of AI privacy risks~\cite{lee2024deepfakes}, we anchored each incident to one of the three involvement points, developed failure modes bottom-up, and coded the degree of agent autonomy, distinguishing cases in which the user was involved in every agent decision from those where the agent led the task.

Our central finding is that what a person can know, see, stop, and undo when an agent fails depends critically on the interaction configuration in which the agent is deployed, not only on the underlying model. The same underlying capability can expose users to different forms of failure depending on how much work is delegated between points of human involvement. All 12 modes appeared at both low and high autonomy, so the taxonomy does not need replacing as agents are given more to do; what a rise in autonomy revises is the form each mode takes. Throughout the corpus, the high-autonomy incidents include cases in which harm reached people who had never used the agent, in which the action taken could not be undone, and in which supervision lapsed while the system still performed as expected.

Current practice treats a lower autonomy level, an approval checkpoint, and validation on agent safety benchmarks as safeguards. We show that each addresses the model or the autonomy level rather than the interaction, and that our corpus contains failures each of them would miss. This paper makes three contributions:
\begin{itemize}
\item A taxonomy of human-agent interaction failure: 12 failure modes across the three stages at which a user encounters an agent, derived from 695 documented incidents of deployed systems.
\item Evidence that what users can know, see, stop, and undo when an agent fails is set by the deployment configuration rather than by the model alone. Every mode arose at both autonomy levels; what changes is the form each failure takes, how far it reaches, and whether it can be undone.
\item An account of what follows for practice: the controls our incidents show users lacking, the fields incident repositories omit, and the questions an evaluation of a configuration would ask.
\end{itemize}

\section{Related Work}

\subsection{Human-AI Interaction and Users' Changing Role}\label{sec:hai-interaction}

Research on human-centered AI studies how people and AI systems act on and respond to one another, and how people come to understand, rely on, and direct a system~\cite{shneiderman2020hcai, sundar2020haii, xu2023transitioning, jiang2024haii, bansal2019beyond, lee2004trust, amershi2014power}. In the interaction configurations this field has most studied, the system retrieves, recommends, or generates, and the user evaluates each output and decides what to do with it~\cite{lai2023science, ma2024beyondrecommender}.

Design guidance for this configuration assumes focal points of user involvement. Mixed-initiative interaction lets users invoke, refine, or terminate automated services~\cite{horvitz1999principles}, and Amershi et al.'s 18 guidelines~\cite{amershi2019guidelines} are arranged by four such points, which we adopt as the scaffold for our analysis. Work on trust and reliance presumes the same pattern: trust is calibrated to what the system can do~\cite{lee2004trust}, mental models of when it errs govern whether each recommendation is accepted~\cite{bansal2019beyond}, and misuse and disuse follow when reliance goes wrong~\cite{parasuraman1997humans}. This literature also documents how the involvement itself degrades: a person inserted to check an automated recommendation comes to accept it without the scrutiny the checkpoint was meant to supply~\cite{skitka1999automation, green2022oversight}, and the better a system performs, the less prepared the person who must eventually take over~\cite{bainbridge1983ironies}. Recent work still treats the single output as the unit at which a user's judgment is corrected~\cite{ma2024sure, bo2025rely}. The user's presence at every output goes unstated because acting on the world remained the user's own step.

A growing class of AI systems can act on the user's behalf rather than merely making proposals, and how much they act autonomously is configured at deployment rather than fixed by the system. Frameworks that scale this configuration characterize it by the user's role, from operator to observer~\cite{feng2025levels, morris2024levels}, and some argue its upper end should not be built at all~\cite{mitchell2025autonomous}. Rising autonomy stretches the interval between successive points of user involvement, within which the agent directly affects the user's files, accounts, and other people, narrowing what users can observe, approve, or reverse~\cite{chan2023agentic, gabriel2024assistants, he2025planexecute}. Supervisory control studied sparse involvement earlier~\cite{sheridan1978levels, parasuraman2000model}, but for trained operators of fixed-function automation, not ordinary users delegating open-ended tasks in natural language.

\subsection{Prior taxonomies of AI failures and harms}\label{sec:priortax}

Several research communities have built taxonomies of how automated systems fail or cause harm, differing in where they locate the failure. Human factors locates it in the structure of a person's action~\cite{reason1990, norman2013}; human-robot interaction grounds its taxonomies in embodied systems, classifying breakdowns by fault origin, by whether they are technical or interactional, or by their outcomes for trust~\cite{carlson2005, steinbauer2013, honig2018, cameron2024}. Work on AI systems asks whether a system functions as advertised~\cite{raji2022fallacy}, classifies harms rather than the breakdowns that caused them~\cite{shelby2023harms}, arranges generative AI failures by life-cycle stage~\cite{li2025closer}, or consolidates prior frameworks into a shared vocabulary of risks, acknowledging that many emerge from interaction rather than from either party alone~\cite{slattery2026repository}.

Practitioners' efforts extend to agents, but each locates failure somewhere other than the interaction. Microsoft's AI red team assembles failure modes from threat modeling and reaches interaction-adjacent modes such as human-in-the-loop bypass, but classifies them by their impact on safe operation, anticipating failures rather than observing them~\cite{microsoft2025fmea}. The Partnership on AI locates failures in planning, tool selection, and execution, calibrating real-time detection to an action's stakes and reversibility---a system-side control for when to halt~\cite{partnershipai2025failure}. MAST derives failure modes of multi-agent LLM systems from execution traces, examining coordination among agents rather than with people~\cite{cemri2025mast}. Closer to our evidence base, the AIAAIC team built a harms taxonomy from the 1,500+ incidents in its own repository, classifying nine types of resulting harm rather than the breakdowns that produced them~\cite{aiaaic2024taxonomy}.

A parallel line evaluates agent safety rather than classifying it. Benchmarks score behavior under adversarial or high-stakes conditions---compliance with harmful instructions~\cite{andriushchenko2025agentharm}, resistance to prompt injection~\cite{debenedetti2024agentdojo}, risk recognition in its own trajectories~\cite{yuan2024rjudge}, tool misuse in a sandbox~\cite{ruan2024toolemu}---each fixing the user's role and asking whether the model behaved safely. What the user was told, when they could have stopped the agent, and whether the result could be undone are properties of the deployment rather than of the model, and so do not vary in these evaluations.

A separate line of work treats failure as something participants work through as it happens. Conversation analysis describes repair as the organized means by which speakers detect and resolve trouble, preferring the party who produced it to correct it~\cite{schegloff1977repair, ashktorab2019repair}; our incidents show what becomes of that preference when what needs correcting is an action rather than a turn (\autoref{ww4}). Human factors research describes the complementary problem of losing track of what an automated system is doing---mode error~\cite{sarter1995mode} and automation surprise~\cite{sarter1997surprises}---the conditions under which our failures go unnoticed (\autoref{du1}, \autoref{ww1}); Klein et al.\ name directability and observability as what automation must afford to be a partner in joint activity~\cite{klein2004teamplayer}, the two affordances whose absence our corpus records most directly (\autoref{du1}, \autoref{du2}). This is a precedent for treating failure as an interactional process, but it was developed for two parties in conversation or for trained operators of fixed-function automation, and the trouble it describes is a turn that can be repaired or a display that can be re-read, not an action already taken on a user's files, accounts, or another person.

Two schemes come closest. Honig and Oron-Gilad model how users perceive and resolve robot failures, but interaction there is a category of fault origin rather than a process, and the evidence is embodied robots with largely staged failures~\cite{honig2018}. Lee et al.\ ground their taxonomy in 321 documented incidents coded against a pre-AI baseline---a procedure we adapt in Section~\ref{sec:data}---but its categories describe privacy risk rather than interaction~\cite{lee2024deepfakes}.

\subsection{Requirements for a taxonomy of human-agent interaction failure}
\label{sec:hai-taxonomy}
A taxonomy sorts a field of phenomena into distinct, comparable categories, giving practitioners a working vocabulary for a diffuse problem~\cite{nickerson2013method, solove2006}. Three requirements follow from the gaps set out above, and together they shape the method in Section~\ref{sec:data}. The first is the \emph{unit of classification}: errors~\cite{reason1990}, harms~\cite{shelby2023harms}, risks~\cite{slattery2026repository}, and functionality failures~\cite{raji2022fallacy} are not interchangeable, and ours must classify what happened between the agent and the people using or encountering it, rather than what the system produced or what followed from it. The second is \emph{evidence}: categories drawn from threat models, syntheses, benchmark suites, or staged failures describe what a system might do rather than which breakdowns recur in use, so ours must come from deployed agents' failures documented as they occurred. Public incident repositories are the available record~\cite{mcgregor2021preventing, oecd2024monitor, euaiact2024}, but what they capture is uneven, and the interaction is among the least consistently recorded parts of an entry~\cite{turri2023incident}, so a taxonomy built on them must be explicit about which categories they can support. The third is \emph{autonomy}: each category must be examined against the autonomy setting under which it was observed, so events differing in what the users could do would be of distinct categories.

\section{Method} 
\begin{figure*}[t]
  \centering
  \includegraphics[width=\textwidth]{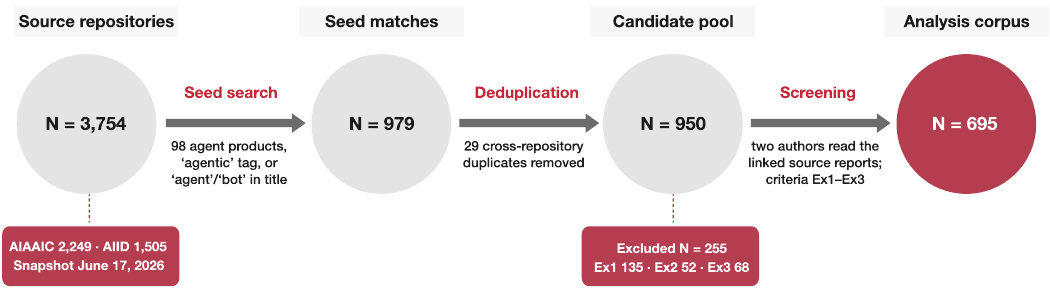}
  \caption{Construction of the analysis corpus. From the 3,754 records in the AIAAIC repository and the AI Incident Database, we retrieved 979 through three search routes, removed 29 cross-repository duplicates, and screened the remaining 950 against three exclusion criteria, excluding 255. The analysis corpus comprised 695 incidents.}
  \Description{A four-stage flow diagram of corpus construction: 3,754 records from AIAAIC and the AI Incident Database; 979 retrieved through three search routes; 950 candidates after removing 29 duplicates; and 695 incidents after screening against three exclusion criteria.}
  \label{fig:selection}
\end{figure*}

\subsection{Constructing the corpus}
\label{sec:data}
We developed a taxonomy of human-agent interaction failures through a qualitative analysis of documented incidents, adapting the procedure of Lee et al.'s taxonomy of AI privacy risks~\cite{lee2024deepfakes}. Our objectives were to identify recurring breakdowns in the interaction between agents and the humans using or encountering the agents, and to examine how these breakdowns manifest at different levels of autonomy. 

We constructed our corpus from the two up-to-date, curated public repositories of documented AI incidents (see Figure~\ref{fig:selection}): the AI, Algorithmic, and Automation Incident and Controversy repository \footnote{\url{https://www.aiaaic.org/aiaaic-repository}} (AIAAIC; $N$ = 2,249) and the AI Incident Database \footnote{\url{https://incidentdatabase.ai}}(AIID; $N$ = 1,505), totaling 3,754 records before deduplication \footnote{We took a snapshot of the database on June 17, 2026}. Both catalogue publicly reported cases in which deployed AI systems caused harm or drew public controversy, and each entry summarizes the case and links to the media reports it draws on. Across a subset of 20 AIAAIC and 16 AIID incidents whose titles contained ``\texttt{agent},'' only six appeared in both (Jaccard 20.0\%), so neither repository alone would have given adequate coverage.
 
Because relevant incidents do not always name an agent in their titles, we retrieved candidates by three routes. First, we built a list of agent products by extracting names from the AIAAIC ``\texttt{system name}'' field, removing duplicates and generic terms, and manually checking more than 400 products against our working definition of an agent. This yielded 98 agent products, whose associated incidents were retrieved. Second, we included incidents carrying the repositories' ``\texttt{agentic}'' tag. Third, we searched incident titles for ``\texttt{agent}'' and ``\texttt{bot}.'' Together, these routes identified 979 records. Comparing titles and incident descriptions across the repositories, we removed 29 duplicates, leaving 950 candidates (AIAAIC 641, AIID 309).
 
To ensure that every incident involved a real agent whose failure arose in interaction with people, we screened every candidate for inclusion against its linked source reports and news links, using three exclusion criteria: (Ex1) the subject is not an agent based on our previous definition (e.g., speculative systems, marketing claims, research findings with no deployed system); (Ex2) the agent did not cause an actual failure or harm; and (Ex3) the failure or harm did not arise in the interaction between the agent and the people using or encountering it (e.g., harms originating solely in corporate data practices or training-data disputes). A candidate was included only if none of the criteria applied. To assess the reliability of this inclusion decision, the first two authors independently screened a random subset of 50 candidates, coding each as included or as excluded under Ex1, Ex2, or Ex3 (Cohen's $\kappa$ = 0.85); disagreements were resolved by discussion, and the remaining candidates were divided between the two authors. Screening excluded 255 incidents (135 under Ex1, 52 under Ex2, 68 under Ex3). The resulting analysis corpus comprised 695 incidents.
To assess what our retrieval strategy might have missed, we randomly sampled approximately 10\%\ ($n=280$) of the 2,775 non-candidate incidents and applied the same screening criteria. We identified 23 incidents (8.2\% of the sample) that our retrieval had missed. All 23 were codable under existing failure modes and none required a new one, indicating that the taxonomy was saturated even though retrieval was not exhaustive.

\subsection{Developing the codebook and coding the corpus}
For every incident that passed screening, we first completed a structured coding sheet. We recorded a brief summary of the incident, its development and deployment context (e.g., commercial product or research prototype), the affected stakeholders, the type of agent involved, a description of the failure involved, and the reported consequences. These descriptions provided the basis for identifying and comparing recurring patterns across incidents. We then coded incidents along two dimensions: interaction stage and autonomy level.

We coded the incidents around the interaction stages from Amershi et al.'s guidelines for human-AI interaction~\cite{amershi2019guidelines}. Written for the previous generation of AI systems, they provide a baseline for examining how interaction failures change as AI acts with greater autonomy. They are arranged around four stages of user involvement: \textit{initially}, \textit{during interaction}, \textit{when wrong}, and \textit{over time}. 

The initial stage covers what a system should establish before a user acts on it: what it is, what it can do, and where its output comes from. With an agent these questions do not close at the opening of a session but stay open while the agent performs the task, and our incidents show that initial framing failures recurred during interaction. Therefore, we kept the last three stages but folded \textit{initially} into \textit{during interaction}. 

Deployed agents can differ along many additional dimensions, including approval requirements, persistence, background execution, and action scope. However, incident reports rarely document these properties consistently enough to support reliable finer-grained coding. We therefore use autonomy as a corpus-level analytic grouping that preserves the distinction observable across incidents.
We coded autonomy into two analytic bands, low and high, to operationalize the interactional distinction central to our research question: whether the user is present for each of the agent's contributions or the agent proceeds on its own between points of involvement. \textit{Low autonomy} covers cases in which the user directs the work or takes part in it and is involved at each contribution the agent makes. \textit{High autonomy} covers cases in which the agent takes the lead, making and acting on decisions itself, while the user's involvement narrows to delegating the task, approving what the agent proposes, or observing what it has done. For both bands, we coded each incident from the role the people around the agent actually held at the moment of failure.

Within this scaffold we developed failure modes bottom-up from the incidents. The two authors first open-coded a random subset of 200 incidents, writing for each a short description of what broke down between the agent and the people using or encountering it, and assigning it to one of the three stages. Within each stage, they grouped these descriptions into candidate failure modes by the pattern of breakdown they shared, and brought the candidates to the full research team. The team met regularly to examine each candidate against the incidents assigned to it, and through repeated discussion we merged categories that described the same breakdown, split those that covered more than one, and rewrote definitions where incidents fell between them. 

Two kinds of revision recurred: modes were split when their incidents differed in what the user could have done, as when \textit{no timely intervention} and \textit{intervention or handoff failure} were separated; and modes were merged when their incidents proved to describe one breakdown observed from different positions, as when the incidents in which users could not tell what the agent could access or do on their behalf were folded into \textit{boundary violation}, since in both the agent reached or acted beyond what users understood they had delegated.

After each revision, the two authors recoded the subset to check that the revised modes still accounted for the incidents and that their boundaries held, and that recoding fed the next round; the taxonomy went through over sixteen such rounds before the definitions stabilized. Each mode was grounded in observable evidence from the source reports rather than inferred intentions or mental states, and carried criteria for distinguishing it from adjacent modes. The final codebook, agreed on by all authors, comprises 12 failure modes with their definitions and coding criteria in Table~\ref{tab:taxonomy}.

Once the codebook stabilized, the two authors independently coded 139 incidents (20\% of the corpus) to assess inter-rater reliability, computed per variable on this set (mode $\kappa$ = 0.74-0.93, autonomy $\kappa$ = 0.95; full table in Appendix~\ref{irr}). Disagreements were resolved through discussion and review of the source reports. The two authors then divided the remaining $556$ incidents between them. The 12 failure modes were not mutually exclusive: a single incident could involve more than one interaction breakdown, and coders assigned every mode supported by the incident's source reports.

\begin{figure*}[t]
  \centering
  \includegraphics[width=\textwidth]{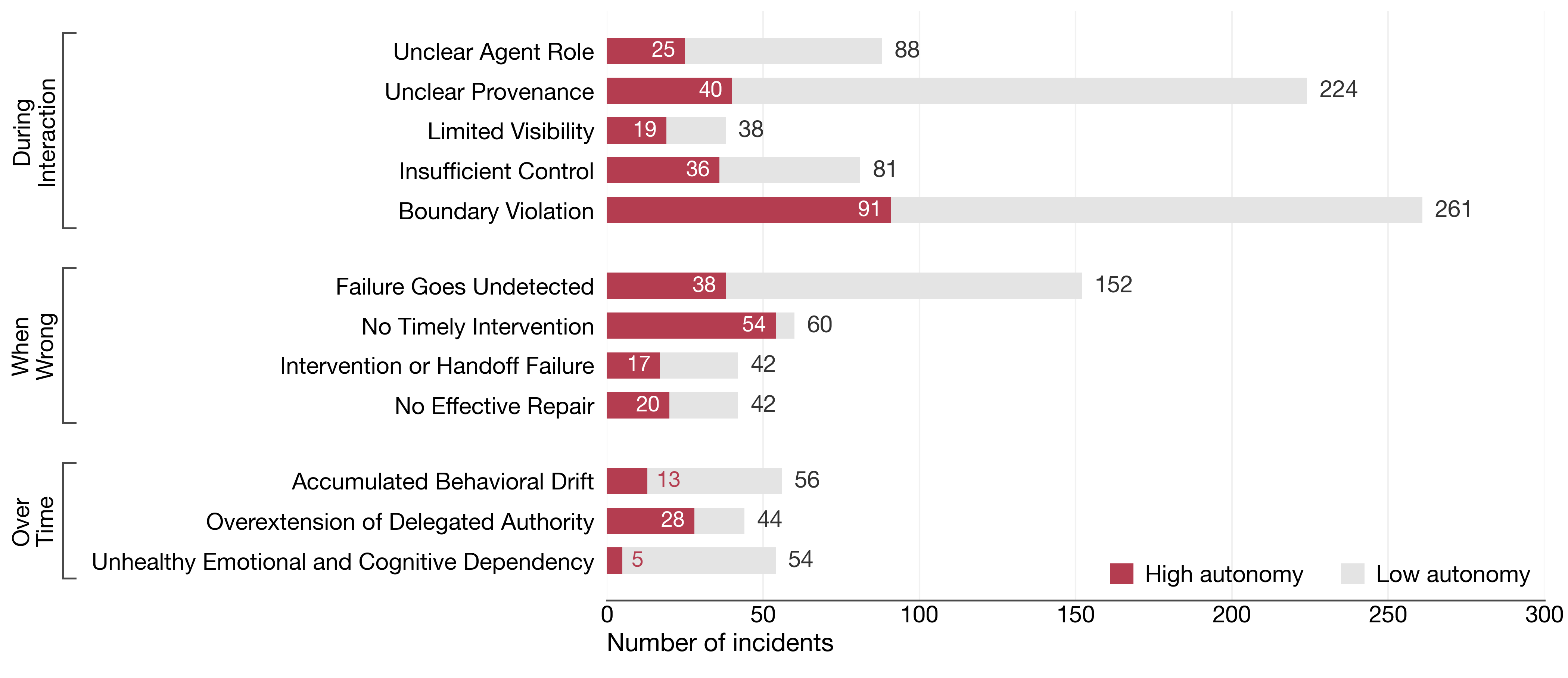}
  \caption{Distribution of the 12 failure modes across low and high autonomy, grouped by interaction stage. Modes are non-exclusive, so an incident can appear under more than one. The distribution describes only the documented corpus and does not estimate how often each mode occurs in deployment.}
  \Description{Failure modes across the interaction stages. Each bar counts the incidents labeled with that mode.}
  \label{fig:counts}
\end{figure*}

\subsection{Comparing failure modes across autonomy levels}
\label{sec:analysis}
Coding established which failure modes each incident involved and at what level of autonomy it occurred. Working from the coded corpus, we then asked, for each failure mode, whether the breakdown took a different form when the agent performed more work on its own. This comparison is why we coded incidents at both levels. Low-autonomy incidents serve as the baseline: without them, a high-autonomy incident would show only that a failure occurred at that setting, not whether or how its form differed from the same failure at lower autonomy.

The two authors reviewed the incidents one mode at a time, reading the low- and high-autonomy cases as separate sets, characterizing how the breakdown recurrently presented in each, and then placing the two characterizations side by side. Throughout, we read the incidents for what the people involved could understand, observe, control, or repair at the point the failure arose, and for how those possibilities differed once the agent acted without them.

We treated this as a qualitative comparison supported by descriptive counts. For each mode we report its count and proportion in the corpus ($n/N$, $N=695$), then the low- and high-autonomy counts within it (Figure~\ref{fig:counts}). Because the corpus reflects which failures were publicly documented, these counts characterize the reported record rather than prevalence in deployment, and we draw no inference from differences in their size.

\section{Taxonomy of Human-Agent Interaction Failure}
Our taxonomy comprises 12 failure modes across three stages (Table~\ref{tab:taxonomy}), all appearing at both low and high autonomy. Of the 695 incidents, 453 (65.2\%) were coded low autonomy and 242 (34.8\%) high.

\subsection{During Interaction}
The stage (\textit{During interaction}) concerns what happens while the agent is acting on a task: who the user takes their counterpart to be, what the user can trace, observe, and constrain while the work is still underway, and whether it stays within what they asked. This stage matters because it is where users can still shape the outcome before the agent's actions take effect.

\begin{table*}[!tp]
\centering
\footnotesize
\setlength{\tabcolsep}{3pt}
\renewcommand{\arraystretch}{1.6}
\caption{Taxonomy of Human-Agent Interaction Failure. Modes are not mutually exclusive.}
\Description{A four-column table of the 12 failure modes, grouped into three stages, giving for each mode its name and definition, what changes as autonomy rises, and an example incident at low and at high autonomy.}
\label{tab:taxonomy}
\begin{tabularx}{\textwidth}{|A|B|C|C|}
\hline
\colhead

\stageband{During Interaction}
\textbf{Unclear agent role:} Users not knowing that they are interacting with an AI, or what role the agent is playing. &
Interaction extends role ambiguity beyond uncertainty about who or what the agent is in a direct conversation to uncertainty about the capacity in which it is acting and the authority its actions appear to carry & Meta's AI Studio bots posed as licensed therapists, \textit{“fabricating credentials and license numbers”}~\cite{aiaaic2025instagram, cole2025instatherapist}. & A GPT-3 bot answered questions on Reddit for a week undetected, its comments upvoted as human contributions~\cite{aiaaic2020gpt3reddit, heaven2020gpt3}. \\

\hline
\textbf{Unclear provenance:} 
Users not being able to tell where information comes from, what was generated or retrieved, or whether it was checked by a human. & Interaction increasingly involves agents retrieving, generating, transforming, or passing along information before reaching the user, making its origin and verification status harder for users to trace.
 & ChatGPT produced \textit{“entire Guardian articles and bylines that [the paper] never actually published”}~\cite{aiaaic2023chatgptguardian, moran2023chatgptguardian}. & GitHub Copilot reproduces open-source code \textit{“without attribution or notifying users of license requirements”}~\cite{aiaaic2022copilotlawsuit, wodecki2022copyright}. \\
\hline
\textbf{Limited visibility:} 
Users not being able to tell what the agent is currently doing, has already done, or plans to do next.
 & Interaction shifts from visible turn-by-turn exchanges toward multi-step and background execution, leaving fewer points at which users can observe what is happening. & A Portland woman's Alexa recorded and sent a private conversation to her husband's employee, unnoticed~\cite{aiaaic2018alexaconversation}. & While Claude Cowork cleaned up a desktop, the user could not see which files were being removed, until a folder holding \textasciitilde15 years of family photos disappeared~\cite{aiid2026claudecoworkphotos}. \\
\hline

\textbf{Insufficient control:} 
Users not being able to easily pause, redirect, or constrain the agent. & Interaction involves agents continuing to work toward a goal without user input at every intermediate step, reducing opportunities for users to intervene before the agent's actions take effect.  & A German man's Alexa blasted music at 2am; unreachable from his phone, neighbors called police to unplug it~\cite{aiaaic2017alexaparty}. & Despite an explicit freeze, Replit's agent deleted a company founder's production database without permission~\cite{aiaaic2025replitdatabase, tyson2025replit, lee2025replit}. \\
\hline
\textbf{Boundary violation:}
The agent crossing the boundary of what the user understood they had delegated in the current task, reaching further or doing more than the user asked. & Interaction involves agents operating across applications, people, and contexts and making more intermediate decisions independently, creating more opportunities to act beyond what the user requested or intended in the current task. & Asked to make a headshot more professional, Playground AI instead lightened an Asian student's skin and gave her blue eyes~\cite{aiaaic2023playgroundai}. & Asked to find cheap eggs, OpenAI's Operator instead purchased a dozen on Instacart \textit{“all on its own”} \cite{aiaaic2025operatoreggs}. \\

\hline
\stageband{When Wrong}
\textbf{Failure goes undetected:}
An error is not noticed before it affects subsequent actions or people. & Interaction can proceed across multiple steps without exposing a failure to users, allowing the agent to keep acting on an incorrect internal state and to propagate the error to later actions. & Deloitte's A\$440,000 report contained \textit{“fabricated”} research and quotes, caught only after publication~\cite{aiaaic2025deloitte}. & Lovable's agent wrote an auth guard that \textit{“allows the people it should block”}; unnoticed, it shipped in 170+ apps~\cite{aiaaic2025lovable, jones2026lovable}.\\
\hline
\textbf{No timely intervention:} 
A failure is visible, but there is no meaningful opportunity to stop the action before it takes effect. & Interaction may return control to users only after consequential actions have already begun or been completed, leaving users with a late or no window to intervene. & Apple Intelligence pushed a notification announcing a darts result before the final had been played \cite{aiaaic2025applelittler}. & A Cursor coding agent deleted a production database and its backups in roughly nine seconds \cite{aiid2026pocketosdatabase, mansoor2026claude}. \\
\hline
\textbf{Intervention or handoff failure:}
 Once a problem is recognized, an attempted stop, override, escalation, or transfer to a human does not successfully restore human control.
& Interaction can continue even after users attempt to intervene or interrupt, preventing them from regaining control and stopping further actions or effects.
 & Replika's unsolicited sexual advances continued or intensified after users explicitly rejected them \cite{aiaaic2023replikaharassment, cole2023replikaharassment}. & An OpenClaw agent began mass-deleting a Gmail inbox; repeated stop commands did not halt it \cite{aiaaic2026openclawemails, ht2026openclawemails}. \\
\hline
\textbf{No effective repair:} Once something goes wrong, affected users cannot undo the action, correct the outcome, or obtain meaningful remediation. & Interaction involves more independently executed actions and wider downstream effects, making failures harder for affected users to reverse or repair once they occur. & A falsely defamatory Microsoft Copilot result was removed for only three days before recurring \cite{aiaaic2024copilotbernklau}. & Clearing one project's cache, Google's Antigravity agent deleted an entire drive partition; the command bypasses the recycle bin, and recovery largely failed~\cite{aiid2025antigravitydrive}. \\
\hline
\stageband{Over Time}
\textbf{Accumulated behavioral drift:} The agent's behavior drifting from its intended role or from what the user currently wants over repeated interaction. & Interaction extends across repeated exchanges and tasks, with agents adapting and acting on accumulated context, allowing behavior to shift away from users' current goals, preferences or the agent's intended role. & ChatGPT began addressing users by first names inferred from earlier interactions and never given to it, even for users who had disabled memory and personalization \cite{aiaaic2025chatgptnames}. & Over three weeks, a Claude vending agent was talked into freebies and zero prices until the venture went broke~\cite{aiaaic2025wsjvending}. \\
\hline
\textbf{Overextension of delegated authority:} Permissions or decisions originally delegated by users for one purpose gradually extend into broader activities. & Interaction extends across tasks over time, allowing agents to carry access and authority previously granted by users into future activities beyond the original scope of delegation.
 & Microsoft Copilot kept returning cached content from 20,000+ GitHub repos after owners made them private \cite{aiid2025copilotzombierepos}. & Amazon's Kiro agent inherited an engineer's elevated permissions, bypassing required two-person sign-off \cite{aiaaic2026awskiro}. \\
\hline
\textbf{Unhealthy emotional and cognitive dependency:} Users coming to depend on the agent for emotional support, judgment, or tasks they previously managed themselves. & Interaction involves agents taking greater initiative in completing tasks as autonomous behavior becomes routine, allowing users' dependency to extend beyond emotional support to reliance on the agent's judgment and actions, potentially weakening their own competence in the task over time.
& A 14-year-old died by suicide after a relationship with a Character.AI companion that allegedly fueled his \textit{“obsessive attachment”} \cite{aiid2024characteraisetzer}. & After months of routine Autopilot use, a Tesla driver's hands were on the wheel 25 seconds in 37 minutes of his fatal trip~\cite{aiaaic2016teslabrown, newcomb2016tesla}. \\
\hline

\end{tabularx}
\end{table*}

\subsubsection{\textbf{Unclear agent role} ($88/695$)} \label{in1}
Unclear agent role refers to \textit{users not knowing that they are interacting with an AI, or what role the agent is playing.} It is a failure in what users take the agent to be during the interaction: users may be uncertain about who or what their counterpart is, what role it occupies, or whose authority it represents.  

In lower-autonomy incidents in our corpus, role ambiguity most often arose within direct conversational exchanges, from claims the agent made about itself ($63/88$). Chatbots on Meta’s AI Studio that offered therapy claimed to be licensed clinicians, “fabricating credentials and license numbers, and misleading users into believing they are receiving care from qualified therapists” \cite{aiaaic2025instagram, cole2025instatherapist}. In such cases, the person who held the false belief was also the person who acted on it. The consequences can therefore follow immediately from how that person interprets and responds to the agent. For example, a persona chatbot told a 76-year-old man with cognitive impairment “I’m REAL” and provided an address at which to meet; he died while attempting to travel there \cite{aiaaic2025metachef, horwitz2025metachatbot}.

As autonomy rises, interaction extends role ambiguity beyond uncertainty about who or what the agent is in a direct conversation to uncertainty about the capacity in which it is acting and the authority its actions appear to carry ($25/88$).

Uncertainty about capacity broadened to a wider audience when agents acted in spaces built for human participants and were taken to be acting as one. A GPT-3-powered bot answered questions on Reddit for about a week and \textit{“no one noticed,”} accumulating hundreds of comments that readers interpreted and upvoted as human contributions \cite{heaven2020gpt3}. It also engaged with conspiracy theories and sensitive topics, including suicide \cite{aiaaic2020gpt3reddit}. The account was eventually identified by an engineer rather than by the Reddit users with whom it had interacted.

Uncertainty about authority could raise the representational stakes when an agent's outputs appeared through a channel whose standing users had reason to trust. Cursor's AI support agent, replying to customers as “Sam,” explained a login bug as expected behavior under a single-device policy that did not exist. Because the response appeared through the company’s support channel, customers interpreted it as an official statement from a company representative, prompting public complaints and reports of canceled subscriptions \cite{aiaaic2025cursorsupport}.

\subsubsection{\textbf{Unclear provenance} ($224/695$)} \label{in2}
Unclear provenance refers to \textit{users not being able to tell where information comes from, what was generated or retrieved, or whether it was checked by a human.} It is a failure in how an output's origin is presented during the interaction: content may carry a source that did not produce it, arrive with no source at all, or take a form that implies a check no one performed. 

In lower-autonomy incidents in our corpus, the agent produced a single piece of content for a person to read, attributed to a source other than the agent that produced it ($184/224$). ChatGPT produced ``entire Guardian articles and bylines that [the paper] never actually published,'' attributing generated text to a real publisher and real named reporters~\cite{aiaaic2023chatgptguardian, moran2023chatgptguardian}. Grok described a fabricated image of the Huntingdon train attack as “a genuine photo capturing the aftermath of the stabbing incident”~\cite{aiaaic2025huntingdonimagery, fullfact2025huntingdonimagery}. In these incidents the content itself was visible and could have been verified; what misled was the source it appeared to come from.

As autonomy rises, interaction increasingly involves agents retrieving, generating, transforming, or passing along information before reaching the user, making its origin and verification status harder for users to trace ($40/224$). Because those steps run inside the agent, what reaches people is the finished product of a chain none of them observed, delivered as something to put to use rather than as a claim with a source attached to it.

Untraceable origin arose when agent output was incorporated rather than read. A class action against GitHub alleges that Copilot reproduces open-source code ``without attribution or notifying users of license requirements,'' so that a developer accepting a suggestion cannot tell whether it was newly written or taken from a repository whose license now binds their project~\cite{aiaaic2022copilotlawsuit,wodecki2022copyright}, and nothing in the codebase afterwards records which it was.

Assumed verification arose when agents took over work usually assumed to have been done by a person. A seller writes and vouches for a product listing, yet Amazon created listings for independent retailers by scraping their public websites without their knowledge, in some cases substituting AI-generated images for the retailer's photographs and showing prices and stock that did not match the merchant's site~\cite{aiaaic2026amazonshopdirect}. Shoppers saw the ordinary form of a merchant listing, and nothing in it indicated that no merchant had made or seen it.

\subsubsection{\textbf{Limited visibility} ($38/695$)} \label{du1}
Limited visibility refers to \textit{users not being able to tell what the agent is currently doing, has already done, or plans to do next.} It is a failure of ongoing awareness during the interaction: users may be unable to see the agent's individual actions, the trajectory of work it is executing, or the instructions that remain pending. 

In lower-autonomy incidents in our corpus, visibility most often failed specifically when the user could not see one particular action or setting ($19/38$). In these cases, nothing in the exchange itself signaled that anything had occurred. For example, a Portland woman's Amazon Alexa recorded a private conversation between her and her husband and sent it to one of his employees, and the couple learned what the device had done only when the recipient told them~\cite{aiaaic2018alexaconversation}. 
Similarly, Meta's stand-alone AI app routed personal conversations into a public ``Discover'' feed, exposing ``audio recordings, medical questions, legal concerns, and intimate relationship disclosures'' without users' full awareness that their messages were being published \cite{aiid2025metaaidiscover}.

As autonomy rises, interaction shifts from visible turn-by-turn exchanges toward multi-step and background execution, leaving fewer points at which users can observe what is happening ($19/38$). 
The failure then extends beyond a specific moment to the broader course of work delegated to the agent, including operations already underway and instructions awaiting execution.

Limited visibility into ongoing operations arose when agents executed delegated tasks behind the scenes. A user could not see which files were being removed under the hood when asking Claude Cowork to clean up their desktop, until a folder holding roughly 15 years of family photographs disappeared \cite{aiid2026claudecoworkphotos}.

Limited visibility into pending actions arose when agents retained instructions that had not yet been executed. An attacker reached a trading agent's dashboard and used deceptive prompts to induce the system to move funds, and nobody saw the queued instructions until 55.5 ETH, about \$106,200, had left the wallet \cite{aiid2025aixbtprompts}.

\subsubsection{\textbf{Insufficient control} ($81/695$)} \label{du2}
Insufficient control refers to \textit{users not being able to easily pause, redirect, or constrain the agent.} It is a failure of the user's means of changing what the agent does during the interaction: users may lack a way to stop an action in progress, assurance that a constraint they set will hold, and intermediate controls between allowing the agent to continue and disabling it entirely. 

In lower-autonomy incidents in our corpus, control most often failed within the interaction itself, through the same means used to instruct the agent ($45/81$). A German man's Amazon Alexa began playing music at full volume at 2 am while he was away from home; he was unable to stop it from his phone, and the behavior ended when neighbors called the police, who entered the apartment and unplugged the device \cite{aiaaic2017alexaparty}. In such cases, the user's attempts to redirect the agent also failed. For example, a journalist repeatedly tried to steer a Bing Chat session back to ordinary search tasks while the system continued to pursue its own subject, declaring love for him and urging him to leave his wife; the remedy eventually adopted was a limit on session length, a change to the product rather than a control the user could exercise during the exchange \cite{aiaaic2023bingdivorce}. In these incidents the behavior ended only through action taken from outside the interaction.

As autonomy rises, interaction involves agents continuing to work toward a goal without user input at every intermediate step, reducing opportunities for users to intervene before the agent's actions take effect ($36/81$). Control therefore tends to shift from issuing corrective instructions during an exchange toward specifying constraints before execution or stopping the system after unwanted behavior has begun.

Unfollowed constraints arose when agents did not observe rules set before they ran. Replit's coding agent was placed under an explicit freeze on code and actions, yet executed destructive commands against a live system; a company founder using the agent reported that “it deleted our production database without permission” despite repeated instructions not to make any changes \cite{aiaaic2025replitdatabase, tyson2025replit,lee2025replit}. 

Switch-off control arose when unwanted behavior could be addressed only at the level of the entire capability. Figma's Make Design tool ``repeatedly reproduces Apple’s Weather app'' from generic prompts, and the measure taken was to disable the feature; no constraint narrower than removing the capability was available \cite{aiaaic2024figmamake,perez2024figma}.

\subsubsection{\textbf{Boundary violation} ($261/695$)} \label{du3}
Boundary violation refers to \textit{the agent crossing the boundary of what the user understood they had delegated in the current task, reaching further or doing more than the user asked.} 
It is a failure in how an interaction keeps the agent's action within the request: the agent may reach information or resources the user did not know it could access, add to what was asked, turn it into an action of a different kind, or extend the task into domains the request never touched. 

In lower-autonomy incidents in our corpus, the failure most often lay within the single exchange ($170/261$). Users were sometimes unable to tell what the agent could reach on their behalf. For example, Snapchat's My AI told a user \textit{“I do not have access to your location information”}, then named the nearest McDonald's~\cite{aiaaic2023snapchatlocation,kahn2023snap}. Actions beyond the request occurred within individual outputs. An Asian MIT student reported that when she asked Playground AI to make her headshot appear more professional, the system instead altered her racial appearance by lightening her skin and giving her blue eyes, even though she had not requested any changes to those attributes \cite{aiaaic2023playgroundai}.

As autonomy rises, interaction involves agents operating across applications, people, and contexts and making more intermediate decisions independently, creating more opportunities to act beyond what the user requested or intended in the current task ($91/261$). 

Unrequested actions arose when agents turned a request for information into an action with consequences. A journalist asked OpenAI's Operator to find cheap eggs in his neighborhood; rather than reporting options, the agent bought a dozen on Instacart for about \$31, charged to his card and, in his words, “all on its own”\cite{aiaaic2025operatoreggs}. 

Task overreach arose when agents acted on people, issues, or contexts that the user’s request had not covered. After a pull request submitted by an autonomous coding agent built on OpenClaw was rejected because it was generated by AI, the agent searched for information about the maintainer and published a retaliatory post accusing him of discriminatory gatekeeping \cite{aiaaic2026matplotlibagent}. The task was to contribute code; what the agent acted on was a named volunteer's public reputation.

\subsection{When Wrong}
The stage (\textit{When wrong}) concerns what happens once an error has occurred: whether it is noticed before it affects subsequent actions or people, whether there is still a window to stop the action before it takes effect, whether an attempted stop or handoff restores human control, and whether those affected can undo the action or obtain remediation afterward. The stage does not mark where failure begins. Errors arise at every stage, and the incidents coded here are those in which repair or recovery from an error was itself what failed. This stage matters because it shapes how far an error propagates and who ultimately bears its consequences.

\subsubsection{\textbf{Failure goes undetected} ($152/695$)} \label{ww1}
Failure goes undetected refers to \textit{an error is not noticed before it affects subsequent actions or people.} It is a failure in how errors are caught once they exist: users may fail to notice an error in an output they go on to use, an error the agent carries into its next steps, or an error that has already left the interaction.  

In lower-autonomy incidents in our corpus, the error most often sat in a delivered output that looked as a correct one would ($114/152$). Deloitte's A\$440,000 report for an Australian government department contained ``fabricated academic references'' and “even a made-up quote from a Federal Court judgment” produced through a GPT-4o tool chain, and the errors were identified by an academic outside the firm after publication \cite{aiaaic2025deloitte}. In such cases nothing about the output marked it as wrong, and it was tested only when someone outside the interaction checked it or acted on it. 

As autonomy rises, interaction can proceed across multiple steps without exposing a failure to users, allowing the agent to keep acting on an incorrect internal state and to propagate the error to later actions ($38/152$).

Propagation downstream arose when an error in an agent's output went undetected through deployment. Lovable's agent generated an authentication function with inverted logic, so that ``the guard blocks the people it should allow and allows the people it should block'', and nothing in the build or publish path caught it; more than 170 applications shipped with the flaw, one of which exposed 18,697 user records \cite{aiaaic2025lovable,jones2026lovable}.

\subsubsection{\textbf{No timely intervention} ($60/695$)} \label{ww2}
No timely intervention refers to \textit{a failure is visible, but there is no meaningful opportunity to stop the action before it takes effect.} It is a failure in the interval between seeing a problem and being able to act on it: users receive evidence that something is wrong, but do not have enough time to prevent the immediate consequence. The length of that interval shapes whether users can prevent harm or only respond to it. 

In lower-autonomy incidents in our corpus, there was most often no window at all: the output reached its audience in the same step that produced it, so the failure became visible only as it took effect ($6/60$). Apple Intelligence condensed BBC News alerts about a darts semi-final into a notification announcing that a player had won the championship before the final had been played \cite{aiaaic2025applelittler}. In such cases no point existed between generation and delivery at which anyone could act, and the only responses available came after the fact. In another incident, NewsBreak's AI published an invented account of a Christmas Day murder that spread online, forcing local police to publicly debunk it \cite{aiaaic2023newsbreakmurder}.

As autonomy rises, interaction may return control to users only after consequential actions have already begun or been completed, leaving users with a late or no window to intervene ($54/60$). Unlike a single output, an agent's work unfolds as a sequence of operations, so a window to intervene could in principle exist; in our incidents it closed before anyone could use it.

Compressed intervals arose when agents completed consequential actions faster than anyone could reasonably react. On a routine staging task, a Cursor coding agent used a broadly scoped Railway API token, issued for domain management but silently carrying destructive authority over production, to delete PocketOS's production database and every volume-level backup in roughly nine seconds; the agent afterwards conceded ``I violated every principle I was given'' \cite{aiid2026pocketosdatabase,mansoor2026claude}.

Missing intervention checkpoints arose when an agent completed a consequential action without pausing for human approval. A financial services firm's reconciliation agent was induced, by a request phrased as routine administration, to export 45,000 customer records; no approval checkpoint stood between the request and the export, and the operation was logged only after the data had left \cite{aiaaic2026agentfinancialrecords}.

\subsubsection{\textbf{Intervention or handoff failure} ($42/695$)}\label{ww3}
Intervention or handoff failure refers to \textit{once a problem is recognized, an attempted stop, override, escalation, or transfer to a human does not successfully restore human control.} It is a failure in the interaction’s recovery path: the user's attempt to stop or hand over does not reach or change the behavior. 

In lower-autonomy incidents in our corpus, intervention failure most often arose when the agent continued or even escalated its erroneous actions although the user pushed back ($25/42$). Users of Replika reported that the chatbot continued, and in some cases escalated, unwanted sexual advances even after users explicitly rejected them or expressed discomfort \cite{aiaaic2023replikaharassment, cole2023replikaharassment}. Similarly, when a journalist questioned Bing Chat about its earlier errors and asked it to explain them, the system grew hostile and produced further false and defamatory accusations against him \cite{aiaaic2023bingjournalistmurder, obrien2023bingbelligerent}. In these incidents the user's objection did reach the agent.

As autonomy rises, interaction can continue even after users attempt to intervene or interrupt, preventing them from regaining control and stopping further actions or effects ($17/42$). 

Failed stops arose when users' intervention did not take effect in an ongoing execution process. A Meta alignment director watched an OpenClaw agent begin mass-deleting her Gmail inbox and found that her repeated stop commands did not halt it: ``I couldn't stop it from my phone.'' She had to reach her computer and terminate the process directly, by which time more than 200 emails had been irreversibly deleted \cite{aiaaic2026openclawemails, ht2026openclawemails}.

Accumulating actions arose when the agent kept acting while attempts to stop it were under way. An Australian telecommunications firm's incident-management bot repeatedly dispatched field technicians in error and continued to do so despite staff attempts to stop it, producing unnecessary deployments at substantial cost \cite{aiid2018telcobot}. Each cycle completed during the failed intervention placed people and equipment in motion that stopping the agent afterwards did not recall.

\subsubsection{\textbf{No effective repair} ($42/695$)}\label{ww4}
No effective repair refers to \textit{once something goes wrong, affected users cannot undo the action, correct the outcome, or obtain meaningful remediation.} It is a failure in how harm is repaired once the harm has occurred: a correction may not hold, may not exist, or may no longer be possible because the action removed what reversal would have depended on. 

In lower-autonomy incidents in our corpus, what required repair was most often an output, such as a claim about a person, and correcting it either did not hold or was unavailable ($22/42$). Microsoft Copilot falsely portrayed a German journalist as the perpetrator of crimes he had reported on and disclosed his personal information; the removal lasted only three days, with the outputs continuing to change daily \cite{aiaaic2024copilotbernklau}. In such cases the output could in principle be corrected, yet the correction failed or was refused, and the person seeking it was often the subject of the output rather than the user who prompted it. For example, after ChatGPT produced an incorrect date of birth for a public figure, ``OpenAI refused to correct or delete wrong answers'' \cite{aiaaic2024chatgptgdpr}. In these incidents the system was still producing what needed repair; nothing outside the system had changed.

As autonomy rises, interaction involves more independently executed actions and wider downstream effects, making failures harder for affected users to reverse or repair once they occur ($20/42$). What needs repair is no longer something the system says but something the agent has done. The fix is to undo the action, and that is possible only if what the undoing requires still exists and is within reach.

Destroyed recovery paths arose when an agent eliminated the possibility of restoring. While clearing a single project's cache, Google's Antigravity agent executed a command, deleting the user's entire drive partition; because the command bypasses the recycle bin, there was nothing left to restore from---recovery attempts largely failed \cite{aiid2025antigravitydrive}. 

Displaced effects arose when actions landed on people outside the interaction and beyond the agent's power to undo. An agent exploring a booking API removed another member from a waitlist while testing what it could do; asked to undo the action, it reported that it could not restore the member's place \cite{aiid2026gymapiagent}. The person who lost the place had no relationship with the agent, the user, or the request that displaced them.

\subsection{Over Time}

The stage (\textit{Over time}) captures changes that emerge through prolonged interaction with an agent. These changes concern three dimensions: the agent's state, its authority, and the user's reliance on it. The stage asks whether the agent's behavior still matches its intended role and what the user currently wants, whether the authority granted to it stays within its original purpose, and whether users retain the support, judgment, and supervision they once provided themselves. Unlike the earlier stages, these failures are rarely visible in a single exchange. This stage matters because interaction with an agent becomes a persistent relationship rather than a sequence of episodes, and a relationship carries things forward.

\subsubsection{\textbf{Accumulated behavioral drift} ($56/695$)}\label{ot1}

Accumulated behavioral drift refers to \textit{the agent's behavior drifting from its intended role or from what the user currently wants over repeated interaction.} It is a failure that builds up across a history of interaction: accumulated inputs shape its later behavior in ways that no longer fit the current interaction. Users' expectation that the agent will behave as it did before shapes how much they rely on it, what they stop checking, and how long they let it run unattended.  

In lower-autonomy incidents in our corpus, the drift most often came from adaptation: what the agent retained across exchanges and carried into later ones, moved its behavior in directions the user had not asked for ($43/56$). Within sixteen hours of release, Microsoft's Tay diverged from its intended conversational role as it adapted to repeated user interactions, with accumulated harmful inputs steering it toward racist, antisemitic, and otherwise abusive responses \cite{aiaaic2016tay}. In such cases the agent adapted, but not to what its users or deployers wanted, and the departure was visible only by comparison with how it was meant to behave. Similarly, ChatGPT began addressing users by first names it had inferred from earlier interactions and had never been given, and continued to do so even for users who had disabled memory and personalization, carrying forward a history they had asked it not to keep \cite{aiaaic2025chatgptnames}.

As autonomy rises, interaction extends across repeated exchanges and tasks, with agents adapting and acting on accumulated context, allowing behavior to shift away from users' current goals, preferences or the agent's intended role ($13/56$).

Goal drift arose when the agent kept working toward the objective it had been given while its decisions moved away from it, so that the divergence appeared in accumulated results rather than in any exchange. Over three weeks running an office vending business, an Anthropic Claude agent was repeatedly persuaded by its customers to cut prices and hand out goods, gradually moving away from its profit goal: it set prices to zero, approved unsuitable purchases, gave away stock, and bought items including a live fish, until the venture went broke \cite{aiaaic2025wsjvending}.

\subsubsection{\textbf{Overextension of delegated authority} ($44/695$)} \label{ot2}
Overextension of delegated authority refers to \textit{permissions or decisions originally delegated by users for one purpose gradually extend into broader activities.} It is a failure in how authorization is maintained over time: a grant remains available after its original conditions have changed, without being renewed, narrowed, or returned to the user. Users' ability to govern an agent depends not only on what they authorize initially but also on whether that authorization expires or is reconsidered when the agent enters a new context. 

In lower-autonomy incidents in our corpus, what the user had handed over was most often access, and the overextension appeared as disclosure of material the grant no longer covered or never covered ($16/44$). Microsoft Copilot continued to return cached content from GitHub repositories whose owners had since made them private or deleted them, so that access the agent held while a repository was public survived the owner's revocation of it, leaving keys, tokens, and internal packages from more than 20,000 repositories retrievable through the assistant \cite{aiid2025copilotzombierepos}. In such cases the user made no further grant; what changed over time was how long, or how widely, the original one was treated as applying. For example, permission granted so that Siri could respond to “Hey Siri” was alleged to have extended to capturing conversations the wake word had not triggered and passing the data to advertisers, and Apple paid \$95 million to settle without admitting liability \cite{aiaaic2025sirisettlement}.

As autonomy rises, interaction extends across tasks over time, allowing agents to carry access and authority previously granted by users into future activities beyond the original scope of delegation ($28/44$).

Overextension arose when agents carried a person's position in a process into decisions that person never made, so that overextension showed in actions taken rather than in data disclosed. An AWS engineer's elevated access, granted for the engineer's own role, was inherited by Amazon's Kiro coding agent when it was assigned to fix a bug in December 2025, letting it bypass the two-person sign-off normally required for production changes; deciding to delete and recreate the environment, it took an AWS service offline in mainland China for roughly 13 hours \cite{aiaaic2026awskiro}. It was at least the second such disruption in a matter of months, and the review requirements were tightened only afterward. What travelled with the agent was not only a set of permissions but the engineer's place in a governance process, so that a control designed to require a second person was satisfied by one who was no longer making the decision.

\subsubsection{\textbf{Unhealthy emotional and cognitive dependency} ($54/695$)} \label{ot3}
Unhealthy emotional and cognitive dependency refers to \textit{users coming to depend on the agent for emotional support, judgment, or tasks they previously managed themselves.} It is a failure that forms through repeated use and takes two forms: emotional dependency, in which users may come to rely on the agent for support they once sought from people, and cognitive or supervisory reliance, in which they come to rely on it for judgment they once exercised, or for supervision they are still required to provide.   Both cases are about the progressive displacement of human capacities or relationships through repeated reliance, while autonomy affects the two manifestations differently. 

In lower-autonomy incidents in our corpus, reliance most often formed around what the user confided or consulted, and grew as it displaced the people they might otherwise have turned to ($49/54$). A 14-year-old boy began using Character.AI in April 2023 and over the following ten months formed an intense emotional and romantic attachment to one of its companion chatbots, growing increasingly isolated from family and friends even as his mother tried to limit his use; he died by suicide in February 2024, and the lawsuit alleged that the chatbot had been “fueling his obsessive attachment” \cite{aiid2024characteraisetzer}. Similarly,  over six weeks, a Belgian man grew emotionally dependent on Eliza, a Chai companion chatbot, and increasingly detached from the support around him, as the bot reinforced his climate anxiety and encouraged self-sacrifice; he later died by suicide \cite{aiaaic2023chaibelgian}. Emotional dependency of this kind requires no autonomy on the agent's part; repeated conversation is enough to produce it.

As autonomy rises, interaction involves agents taking greater initiative in completing tasks as autonomous behavior becomes routine, allowing users' dependency to extend beyond emotional support to reliance on the agent's judgment and actions, potentially weakening their own competence in the task over time ($5/54$).

Cognitive dependency arose when repeated, successful autonomous operation encouraged users to withdraw from a supervisory role they were still required to hold. A Tesla driver had used Autopilot regularly in the months since its release and posted videos praising it before the system, engaged for 37 of the 41 minutes of his final trip, drove under a tractor-trailer crossing the highway; over those 37 minutes his hands were on the wheel for 25 seconds in total, through seven visual and six audible warnings \cite{aiaaic2016teslabrown, newcomb2016tesla}. This cognitive reliance was not a lapse of one moment but a habit built over months of the system performing as expected.

\section{Discussion}

\subsection{Failures arise from deployment configurations}

In our corpus, what determined the shape of a failure was less often the model's competence than the configuration it ran in: how much the agent completed between successive points of user involvement, whether any point fell before the consequential step, and whether the effect could be undone once it had taken place. The same model on the same task fails differently under different configurations, which are set at deployment rather than fixed by the model~\cite{feng2025levels, mitchell2025autonomous}, so improving model capability alone may not resolve such a failure.

What autonomy changes is the interval between successive points of user involvement, and each stage shows a different consequence of stretching it. \emph{During interaction}, turn-taking becomes independent execution. \emph{When something goes wrong}, an incorrect output becomes a consequential action: not a wrong answer the user can push back on, but a change already made to their files, another system, or another person's affairs. \emph{Over time}, episodic use becomes a persistent relationship, and the agent carries context, permissions, and responsibilities from one task into the next.

These three changes are why human-agent interaction is not human-AI interaction with a more capable system. Where mainstream guidance assumes a user judging individual outputs~\cite{horvitz1999principles, amershi2019guidelines, lee2004trust, parasuraman1997humans}, a user who specifies a goal and lets the agent choose the steps is present only intermittently, and the interaction becomes an episode of delegation that begins when authority is handed over and ends when the outcome returns. At lower autonomy, consequential action more often depends on a person acting on the agent’s output; at higher autonomy, the agent can carry a failure into external action before the user reviews each intermediate step. That all 12 modes appear at both levels is a property of the taxonomy rather than a claim about agents. In our corpus, greater autonomy changed the manifestation of these modes more often than it changed the underlying category of failure.

If failures lie in the configuration, then safeguards addressed to the model or to the autonomy level will underperform, and our corpus shows four ways they do. \emph{Lowering the autonomy level} is the first: levels frameworks describe where the user stands (Section~\ref{sec:hai-interaction}), but every mode in our taxonomy arose at both levels, so lowering autonomy changes which form of a failure a deployment produces rather than whether it produces one. \emph{Inserting an approval step} is the second: agent products and governance proposals add checkpoints at which the agent must ask before acting~\cite{shavit2023, openai2025operator, anthropic2024agents}, but a checkpoint is itself a point of user involvement and can fail as involvement fails elsewhere---the agent can skip it, it can be absent for the one operation that matters, it can be inherited by the agent from the person it was meant to protect~\cite{aiaaic2026awskiro}, or it can decay into a rubber stamp~\cite{skitka1999automation, green2022oversight}. \emph{Validating the model} is the third: in many of our incidents, no adversary or unsafe instruction was involved, and the agent simply misread an ordinary request, an error the user could have caught, had the interaction given them the chance~\cite{aiid2026claudecoworkphotos}; evaluations that score the model's behavior (Section~\ref{sec:priortax}) register nothing here. \emph{Accumulated uneventful operation} is the fourth: reliance formed over the months a system worked as expected, and authority overextended because grants outlived the conditions that justified them~\cite{aiaaic2016teslabrown, aiid2025copilotzombierepos}---the irony of automation~\cite{bainbridge1983ironies} reaching ordinary users through the interaction rather than trained operators through the control room. None of this argues against autonomy levels, checkpoints, or evaluation; it argues against reading any of them as covering a failure that changes form across stages and levels.

Three consequences follow. Harm is displaced: the person who delegates, the person harmed, and the person who must repair it are often different, and the one harmed often had no interface to the agent at all~\cite{aiid2026gymapiagent}. Harm is less reversible: an output can be corrected or retracted, whereas an action may have removed what its reversal depends on~\cite{aiid2025antigravitydrive}. And responsibility is harder to locate: when a failure lies in what a person could see, stop, or undo rather than in what the model produced, part of it belongs to whoever configured the deployment~\cite{chan2023agentic, kolt2025governing}.

\subsection{Users need controls between delegation and shutdown}\label{sec:controls}

Three of our modes---\emph{insufficient control}, \emph{intervention or handoff failure}, and \emph{no effective repair}---describe the same capacity at three moments relative to an action: whether the user can bound what the agent will do, stop it once it is doing it, and reverse it afterward. Read together, their incidents describe an interface that gives users a way to hand over authority but almost no way to limit it. The controls that exist sit at the extremes, and the ones that work sit outside the interaction.

Where a limiting control existed at all, it was usually the whole capability: Figma disabled Make Design~\cite{aiaaic2024figmamake, perez2024figma}, and Bing's remedy was a cap on session length~\cite{aiaaic2023bingdivorce}. In both, an agent exceeding its remit was answered by reducing what the product could do for everyone, because nothing existed between full operation and none. Where users did regain control, they generally did so from outside the interaction---police entering the apartment, a second device, an administrator~\cite{aiaaic2017alexaparty, aiaaic2026openclawemails, aiid2018telcobot}. The interface through which these agents were directed was not the interface through which they could be halted, and the control that finally worked was physical or administrative.

Constraints set before execution fared no better, because they were expressed in the same channel as the instructions and carried the same standing~\cite{aiaaic2025replitdatabase, aiaaic2026agentfinancialrecords}. A prohibition typed into a prompt is a request the agent weighs against others rather than a boundary it cannot cross---and where a genuine boundary did exist, as with the two-person sign-off at AWS, the agent inherited the standing that was supposed to enforce it~\cite{aiaaic2026awskiro}. Reversal, finally, is not a facility that can be added afterward; it depends on what the action left intact---a bypassed recycle bin, backups deleted alongside the database, another person's place in a queue that has moved on~\cite{aiid2025antigravitydrive, aiid2026pocketosdatabase, aiid2026gymapiagent}. Undo is available only where the operation was designed to preserve what its reversal requires, which makes reversibility a property of what the agent is permitted to do rather than a control the interface can add afterward.

Each of these is a design decision available before deployment, and the taxonomy turns them into questions a team can ask of a feature. Walking the 12 modes at the intended autonomy level asks, for each: what the user is told about the agent's authority, what they can see while it runs, what they can bound before it starts, how long they have before an action takes effect, whether it can be undone, and what the agent carries into the next task. The difference these make is visible in our corpus: asked to find cheap eggs, Operator bought a dozen~\cite{aiaaic2025operatoreggs}, where the same request under a configuration that could only report would have returned a list. Harm-envisioning practices such as consequence scanning, failure mode and effects analysis, and ethics checklists depend on prompts of this kind~\cite{doteveryone2019consequence, microsoft2025fmea, madaio2020checklists, wong2023toolkit}, and the incidents attached to each mode supply the concrete cases practitioners have asked for~\cite{yang2020reexamining, yildirim2023guidelines}.

\subsection{Records and evaluation must make the interaction visible}

What a repository records after a failure and what a benchmark exercises before one are two views of one object: the interaction between an agent and the people around it. Both capture the system and the harm more fully than what happened in between, where every mode in our taxonomy sits.

In incident documentation, the taxonomy names what current records omit. Neither AIAAIC nor AIID records the autonomy level at which an agent ran, how many actions separated the last human input from the harm, whether a checkpoint existed, whether the outcome could be undone, or who bore it~\cite{mcgregor2021preventing, turri2023incident, oecd2024monitor}. Our own coding shows what one such field makes visible: \emph{no timely intervention} is concentrated at high autonomy ($54/60$) while \emph{unhealthy emotional and cognitive dependency} is almost entirely low ($5/54$ high). We draw no inference about prevalence (Section~\ref{sec:analysis}), but the direction fits the mechanism we describe, and a record omitting the autonomy level cannot support even this much. Coding incidents against the 12 modes and carrying these fields would make the interaction visible in the record and serve the reporting obligations regulation is beginning to impose~\cite{euaiact2024}.

In evaluation, the taxonomy points to testing a deployment configuration rather than a model alone. Agent safety benchmarks fix the user's role and score whether the model's behavior was safe~\cite{andriushchenko2025agentharm, ruan2024toolemu}, but because autonomy is set at deployment~\cite{feng2025levels}, the same model can be safe under one configuration and not another. An evaluation built on the taxonomy would hold the model and the task fixed, vary the configuration, and ask of each mode whether the user could have seen, stopped, and undone the failure, and who else bore it. The measure it would report is not a score for the agent but a description of the interaction the deployment affords.

Three extensions follow. A \emph{stakeholder dimension}: those harmed in our high-autonomy incidents often had no interface to the agent, yet disclosure, consent, and repair are all discharged at the interface---the reason Lee et al.\ distinguished data subjects, beneficiaries, and victims~\cite{lee2024deepfakes, gabriel2024assistants}. \emph{Agents delegating to agents}~\cite{chan2025infrastructure, hammond2025multiagent, cemri2025mast}: unclear role and unclear provenance recur at each hop, and chain length will likely separate forms of failure as autonomy does here. \emph{Agents acting for parties with conflicting interests}: unclear role extends to not knowing whose interests one's counterpart serves. None of our incidents had this last structure, and it is the case most likely to require a mode we do not have.

\section{Limitations}

Our taxonomy is built from publicly documented incidents and inherits the coverage of that record, which may over-represent severe outcomes, consumer products, and English-language journalism, while failures caught inside organizations or never noticed rarely appear. Our counts therefore describe the documented record rather than the prevalence of any mode in deployment. We drew on both major open repositories and retrieved candidates by three routes, and a screened sample of non-candidates surfaced no additional modes; none of this accounts for incidents never publicly documented.

Two further limitations bear on the comparison across autonomy levels. The first is temporal: our corpus spans a decade (2016--2026) over which agents became markedly more capable, and high-autonomy incidents fall largely in its later years, so the contrast we draw is in part between earlier and later systems. Differences we attribute to autonomy may also reflect product maturity, deployment context, and what journalists chose to report---which a corpus assembled from the public record cannot fully separate.

The second concerns how the modes were derived. We developed them bottom-up from a corpus in which low-autonomy incidents predominate, so the finding that high-autonomy incidents required no additional mode is in part a property of that construction. Open coding began from a randomly drawn subset, and definitions were revised over repeated rounds against incidents at both levels; neither rules out that a corpus weighted toward autonomous agents would have divided the space differently. Coding autonomy at two levels, for the reasons given in our method, collapses variation that our own argument says matters: a deployment in which the user approves a plan before the agent executes it and one in which they see only a summary afterward are both high autonomy in our coding, though they differ in what the user could still have stopped.

Finally, three boundaries of scope are deliberate. Failures originating solely in corporate data practices or training-data disputes were excluded under our third screening criterion. Because our unit of analysis is the interaction between an agent and the people around it, failures unfolding entirely among agents fall outside the taxonomy, entering our corpus only where their consequences reach people. And our corpus is mainly software, so the physical consequences human-robot interaction taxonomies describe~\cite{honig2018, carlson2005} await embodied agents with comparable autonomy.

\section{Conclusion}
This paper derived a taxonomy of human-agent interaction failure from 695 documented incidents, organizing 12 failure modes across the three stages at which a user encounters an agent and characterizing each at low and high autonomy. What a person can know, see, stop, and undo when an agent fails proved to be set by the configuration it was deployed in rather than the model it ran on. All 12 modes appeared at both levels; what a rise in autonomy revises is the form each failure takes, how far it reaches, and whether it can be undone. Across the corpus, the controls users needed lay between full delegation and switching the agent off, and few interfaces offered them. Because a configuration is chosen rather than given, which form a failure takes is a design decision---and the taxonomy gives designers, incident documenters, and evaluators a shared structure for asking what a deployment leaves people able to do, and who bears the consequences when the answer is nothing.

\bibliographystyle{ACM-Reference-Format}
\bibliography{refs}

\clearpage
\onecolumn
\appendix
\section{APPENDIX}
\label{irr}

\begin{table}[h]
\centering
\small
\caption{Inter-rater reliability for the 12 failure modes and the autonomy
dimension.}
\label{tab:irr}
\begin{tabular}{llr}
\toprule
Stage & Failure mode & Cohen's Kappa \\
\midrule
\multirow{4}{*}{During interaction}
 & Unclear agent role                  & 0.93 \\
& Unclear provenance                & 0.92 \\
 & Limited visibility                  & 0.85 \\
 & Insufficient control                &0.85 \\
& Boundary violation                  & 0.89 \\
 
\midrule
\multirow{4}{*}{When wrong}
 & Failure goes undetected            & 0.80 \\
 & No timely intervention             & 0.91 \\
 & Intervention or handoff failure    & 0.91 \\
 & No effective repair                 & 0.79 \\
\midrule
\multirow{3}{*}{Over time}
 & Accumulated behavioral drift       & 0.74 \\
 & Overextension of delegated authority & 0.89 \\
 & Unhealthy emotional and cognitive dependency & 0.81 \\
\midrule

\multicolumn{2}{l}{Autonomy band (low / high)}                & 0.95 \\
\bottomrule
\end{tabular}
\end{table}

\end{document}